\documentclass[a4paper,UKenglish]{lipics}
 
\usepackage{microtype}
\usepackage[utf8]{inputenc}
\usepackage[ruled]{algorithm}
\usepackage{hyperref}\hypersetup{colorlinks=false}
\newtheorem {myexample}{Example} 
\newtheorem {mytheorem}{Proposition} 

\title{\textit{dasasap}, an App for Syllogisms}
\titlerunning{\textit{dasasap}, an App for Syllogisms} 

\author[1]{J. Martín Castro-Manzano}
\author[2]{Verónica Reyes-Meza}
\author[3]{Jorge Medina-Delgadillo}
\affil[1]{Faculty of Philosophy and Humanities, UPAEP\\
  \texttt{josemartin.castro@upaep.mx}}
\affil[2]{Faculty of Psychology, UPAEP\\
  \texttt{veronica.reyes@upaep.mx}}
\affil[3]{Faculty of Philosophy and Humanities, UPAEP\\
  \texttt{jorge.medina@upaep.mx}}
\authorrunning{J.\,M. Castro-Manzano et. al.} 

\Copyright{José Martín Castro-Manzano}



\serieslogo{logo_ttl}
\volumeinfo
  {M. Antonia {Huertas}, Jo\~ao {Marcos}, Mar\'ia {Manzano}, Sophie {Pinchinat}, \\
  Fran\c{c}ois {Schwarzentruber}}
  {5}
  {4th International Conference on Tools for Teaching Logic}
  {1}
  {1}
  {9}
\EventShortName{TTL2015}

\begin{document}

\maketitle

\begin{abstract}
The main goal of this contribution is to introduce a cross-platform application to learn-teach syllogistic. We call this application \textit{dasasap} for \textit{develop all syllogisms as soon as possible}. To introduce this application we show the logical foundations for the game with a system we call $\mathcal{L}_\square$, and its interface developed with LiveCode. 
\end{abstract}

\section{Introduction}
Syllogistic has a remarkable impact in our education and culture: most undergraduate logic courses, books and manuals around the world cover or include a fragment of syllogistic not just to teach logic, but to provide \textit{form\ae\ mentis}. Therefore, whether studying syllogistic is relevant or not is beyond doubt; the problem, if any, has to do with the different treatments used to learn-teach syllogistic. 

The main goal of this contribution is to introduce a cross-platform application to learn-teach syllogistic. We call this application \textit{dasasap} for \textit{develop all syllogisms as soon as possible}. To introduce this application we show the logical foundations for the game with a system we call $\mathcal{L}_\square$ (Section~\ref{sec:1}), and its interface developed with LiveCode (Section~\ref{sec:2}). We close this work with comments about future work (Section~\ref{sec:3}). 

\section{Logical foundations}
\label{sec:1}
\subsection{Syllogisms}
A \textit{categorical proposition} is a proposition with a defined quantity (universal or particular), a subject $\mathsf{S}$, a quality (affirmative or negative), and a predicate $\mathsf{P}$. The adequate combinations of these elements produce four categorical propositions, each one with a name:

\begin{itemize}
\item Proposition $\mathsf{A}$ states $All\ \mathsf{S}\ is\ \mathsf{P}$.
\item Proposition $\mathsf{E}$ states $No\ \mathsf{S}\ is\ \mathsf{P}$.
\item Proposition $\mathsf{I}$ states $Some\ \mathsf{S}\ are\ \mathsf{P}$.
\item Proposition $\mathsf{O}$ states $Some\ \mathsf{S}\ are\ not\ \mathsf{P}$.
\end{itemize}

A \textit{categorical syllogism} is a sequence of three categorical propositions ordered in such a way that the two first propositions are called \textit{premises} (major and minor) and the last one is called \textit{conclusion}. Within the premises there is a term $\mathsf{M}$, known as \textit{middle term}, that works as a \textit{link} that gathers the remaining terms, namely, $\mathsf{S}$ and $\mathsf{P}$:

\begin{myexample}\text{}
\begin{enumerate}
	\item Major premise: \textit{All} $\mathsf{M}$ \textit{is} $\mathsf{P}$.
	\item Minor premise: \textit{All} $\mathsf{S}$ \textit{is} $\mathsf{M}$.
	\item Conclusion: \textit{All} $\mathsf{S}$ \textit{is} $\mathsf{P}$.
\end{enumerate}
\end{myexample}

The previous example can be represented as a string of characters using the terms and proposition names stated previously in an infix notation: $\mathsf{MAP}\mathsf{SAM}\therefore\mathsf{SAP}$, where $\therefore$ indicates that the premises are to the left and the conclusion to the right. According to the position of the middle term $\mathsf{M}$ we can build four \textit{figures} of the syllogism that encode all the valid and only the valid syllogisms. These have been named using mnemonics~\cite{PETEROFSPAIN} (Table~\ref{tab:1}).

\begin{table}[h]
\caption{Valid categorical syllogisms}
\label{tab:1}      
\begin{tabular}{llll}
\hline\noalign{\smallskip}
Figure 1 & Figure 2 & Figure 3 & Figure 4 \\
\noalign{\smallskip}\hline\noalign{\smallskip}
Barbara 		  & Cesare 			& Disamis 		    & Camenes 			\\
$\mathsf{MAP}\mathsf{SAM}\therefore\mathsf{SAP}$ & $\mathsf{PEM}\mathsf{SAM}\therefore\mathsf{SEP}$ & $\mathsf{MIP}\mathsf{MAS}\therefore\mathsf{SIP}$ & $\mathsf{PAM}\mathsf{MES}\therefore\mathsf{SEP}$ \\
Celarent 		  & Camestres 			& Datisi 		    & Dimaris 			\\
$\mathsf{MEP}\mathsf{SAM}\therefore\mathsf{SEP}$ &$\mathsf{PAM}\mathsf{SEM}\therefore\mathsf{SEP}$ & $\mathsf{MAP}\mathsf{MIS}\therefore\mathsf{SIP}$ & $\mathsf{PIM}\mathsf{MAS}\therefore\mathsf{SIP}$ \\
Darii 			  & Festino 			& Bocardo 		    & Fresison 			\\ 
$\mathsf{MAP}\mathsf{SIM}\therefore\mathsf{SIP}$ & $\mathsf{PEM}\mathsf{SIM}\therefore\mathsf{SOP}$ & $\mathsf{MOP}\mathsf{MAS}\therefore\mathsf{SOP}$ & $\mathsf{PEM}\mathsf{MIS}\therefore\mathsf{SOP}$ \\
Ferio 			  & Baroco  			& Ferison 		    &        			\\  
$\mathsf{MEP}\mathsf{SIM}\therefore\mathsf{SOP}$ & $\mathsf{PAM}\mathsf{SOM}\therefore\mathsf{SOP}$ & $\mathsf{MEP}\mathsf{MIS}\therefore\mathsf{SOP}$ & 				\\
\noalign{\smallskip}\hline
\end{tabular}
\end{table} 

\subsection{Jigsaw puzzles}
\begin{figure}[h]
  \includegraphics[width=5cm]{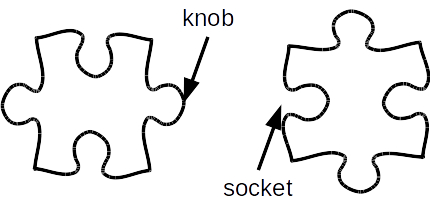}
  \caption{Pieces of a typical jigsaw puzzle}
  \label{fig:0}
\end{figure}

A jigsaw puzzle is a tiling array composed by a finite set of tessellating pieces that require assembly by way of the \textit{interlocking} of knobs and sockets (Figure~\ref{fig:0}). When finished, a jigsaw puzzle mechanically produces a complete picture by the adequate interlocking of all the pieces; when the pieces cannot be interlocked, the jigsaw puzzle cannot be completed, and thus the picture cannot be mechanically produced.

These types puzzles date back as far as Archimedes~\cite{SLOCUM}, so the historical paths of logic and jigsaw puzzles have a long tradition, although the typical pictorial jigsaw puzzles we are familiar with have their roots in Europe when John Spilsbury's first dissected maps in the 1760s primarily for educational purposes~\cite{WILLIAMS}. We suggest that syllogisms can be developed as jigsaw puzzles that require assembly by way of \textit{interlocking}.

\pagebreak

\subsection{The system $\mathcal{L}_{\square}$}
\begin{figure}[h]
  \includegraphics[width=5cm]{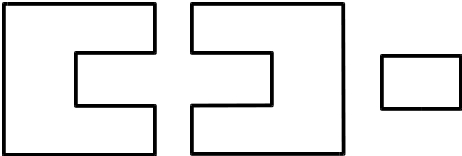}
  \caption{Vocabulary for $\mathcal{L}_{\square}$}
  \label{fig:1}
\end{figure}

$\mathcal{L}_{\square}$ is a system that uses adequate combinations of polygons as jigsaw puzzles in order to decide the (in)validity of categorical syllogisms. Its vocabulary is defined by two elementary diagrams (i.e., pieces), socket and knobs (Figure~\ref{fig:1}). 
 
\begin{figure}[h]
  \includegraphics[width=6cm]{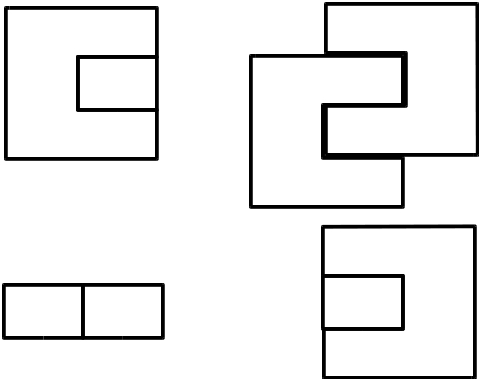}
  \caption{Well formed diagrams for $\mathcal{L}_{\square}$}
  \label{fig:2}
\end{figure}

Syntax is given by a single rule: if a diagram belongs to the vocabulary of $\mathcal{L}_{\square}$, the diagrams depicted in Figure~\ref{fig:2}, and only those, are well formed diagrams. 

\begin{figure}[h]
  \includegraphics[width=5cm]{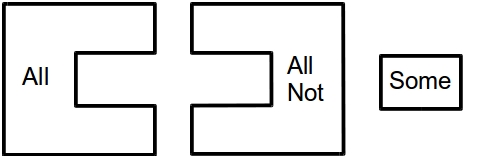}
  \caption{Semantics for $\mathcal{L}_{\square}$}
  \label{fig:3}
\end{figure}

Semantics is given by Figure~\ref{fig:3}. 

\pagebreak

\begin{figure}[h]
  \includegraphics[width=7cm]{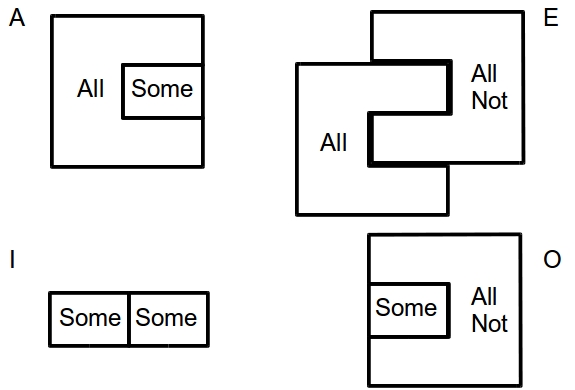}
  \caption{Categorical propositions in $\mathcal{L}_{\square}$}
  \label{fig:4}
\end{figure}

With this system we can represent the categorical propositions as in Figure~\ref{fig:4}. 

\begin{figure}[h]
  \includegraphics[width=7cm]{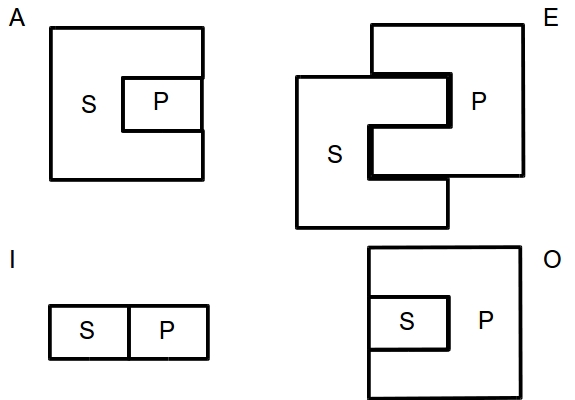}
  \caption{Square of Opposition in $\mathcal{L}_{\square}$}
  \label{fig:5}
\end{figure}

We can arrange this system in a Square of Opposition (Figure~\ref{fig:5}). It should be clear that the only rules preserved in this Square are the rules for \textit{contradiction} between $\mathsf{A}$ ($\mathsf{E}$) and $\mathsf{O}$ ($\mathsf{I}$); the rules for \textit{contraries}, \textit{subalterns}, and \textit{subcontraries} do not work. Thus, this system's Square behaves as a Modern Square of Opposition rather than a Traditional Square. Also, \textit{conversion}, \textit{contraposition}, and \textit{obversion} are valid transformations in this Square.  

\subsection{Decision Algorithm in $\mathcal{L}_{\square}$}
Using a single term, in this case the middle term, we can represent the Identity Principle ($\mathsf{IP}$) in this system: 

\begin{itemize}
\item Proposition $\mathsf{A}$ states $All\ \mathsf{M}\  is\ \mathsf{M}$.
\item Proposition $\mathsf{E}$ states $No\ \mathsf{M}\  is\ \mathsf{M}$.
\item Proposition $\mathsf{I}$ states $Some\ \mathsf{M}\  are\ \mathsf{M}$.
\item Proposition $\mathsf{O}$ states $Some\ \mathsf{M}\  are\ not\ \mathsf{M}$.
\end{itemize}

\begin{figure}[h]
  \includegraphics[width=7cm]{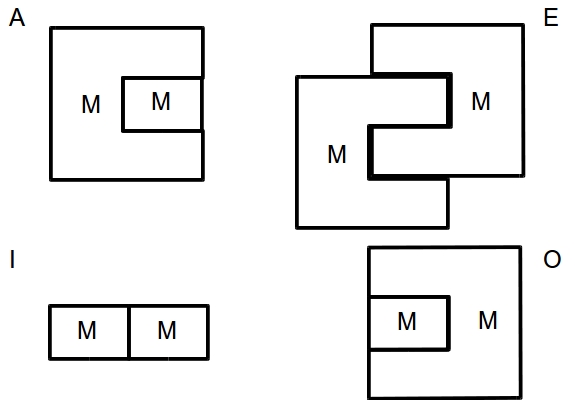}
  \caption{Identity Principle in $\mathcal{L}_{\square}$}
  \label{fig:9}
\end{figure}

Figure~\ref{fig:9} shows the four categorical propositions using the middle term $\mathsf{M}$. We can observe that from these four cases only $\mathsf{A}$ encodes a logical truth, that is, a true statement for any structure, provided it is a weak version of the $\mathsf{IP}$ as in $\forall x(\mathsf{M}x \Rightarrow \exists y\mathsf{M}y)$. 

\pagebreak

Using the $\mathsf{IP}$ we suggest a decision algorithm for syllogistic in $\mathcal{L}_{\square}$ with complexity of $O(N)$ (Algorithm 1).

\begin{algorithm}
  \SetAlgoLined\scriptsize
  \KwIn{syllogism $\sigma$}
  \KwOut{decision $\delta$}
  \For{$\sigma$}{
    \eIf{middle terms of $\sigma$ == $\mathsf{IP}$}{
      $\delta \gets valid$\;
      }{
      $\delta \gets invalid$\;
      }
    }
  \caption{Decision algorithm in $\mathcal{L}_{\square}$}
  \label{alg:1}
\end{algorithm}

Hence, this algorithm not only is time efficient but sound and complete:

\begin{mytheorem}
(Soundness) If $\mathcal{A}(\sigma_{i,j})=valid$ then $\sigma_{i,j}$ is valid.
\end{mytheorem}
\textit{Proof.} Let us denote the application of Algorithm~\ref{alg:1} to a given syllogism $\sigma_{i,j}$ from figure $i\in\lbrace 1,2,3,4 \rbrace$ and row $j\in\lbrace 1,2,3,4 \rbrace$ in Table~\ref{tab:1} by $\mathcal{A}(\sigma_{i,j})$ (thus, for instance, the application of Algorithm~\ref{alg:1} to a \textit{Dimaris} syllogism is $\mathcal{A}(\sigma_{4,2})$). We prove soundness by cases. Since there are four figures, we need to cover each valid syllogism from each figure. Here we only show the first case that covers the syllogisms of figure 1 (\textit{Barbara}, \textit{Celarent}, \textit{Darii}, and \textit{Ferio}) in Section~\ref{subsec:example}. The remaining syllogisms can be proven similarly. Thus, we have that for every $\sigma_{i,j}$, when $\mathcal{A}(\sigma_{i,j})=valid$, $\sigma_{i,j}$ is valid. $\blacksquare$ 

\begin{mytheorem}
(Completeness) If $\sigma_{i,j}$ is valid then $\mathcal{A}(\sigma_{i,j})=valid$.
\end{mytheorem}
\textit{Proof}. We prove this by contradiction. Suppose that for all $i,j$, the syllogism $\sigma_{i,j}$ is valid but for some valid syllogism $\sigma_{k,j}$,  $\mathcal{A}(\sigma_{k,j})=invalid$. Now, we know $\sigma_{1,j}$ is valid and if we apply $\mathcal{A}(\sigma_{1,j})$ we obtain $\mathcal{A}(\sigma_{1,j})=valid$, as we can see in Section~\ref{subsec:example}. So, since all valid syllogisms $\sigma_{n>1,j}$ can be reduced to the valid syllogisms of figure 1 by the rules of equivalence (conversion, contraposition, and obversion), it follows that $\mathcal{A}(\sigma_{n>1,j})=valid$, and thus, for all valid syllogisms $k$, $\mathcal{A}(\sigma_{k,j})=valid$, which contradicts our assumption. $\blacksquare$

\subsection{Example}\label{subsec:example}
\begin{figure}[h]
  \includegraphics[width=14cm]{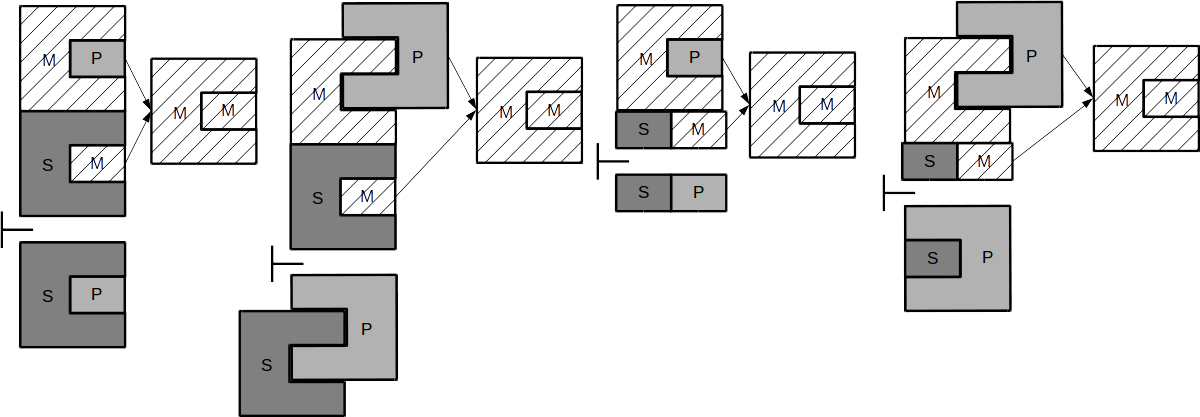}
  \caption{\textit{Barbara}, \textit{Celarent}, \textit{Darii}, \textit{Ferio}. In order to represent and build a categorical syllogism in $\mathcal{L}_{\square}$ we just stack up two well formed diagrams that represent the premises. When the middle terms \textit{interlock} each other forming the $\mathsf{IP}$ in a jigsaw puzzle style (which is a step denoted by the arrows), the inference is valid, thus allowing the terms $\mathsf{S}$ and $\mathsf{P}$ interlock in the third diagram, i.e., the conclusion, which is below the turnstile.}
  \label{fig:10}
\end{figure}

As an example of how $\mathcal{L}_{\square}$ works we prove the four valid syllogisms of figure 1 (Figure~\ref{fig:10}), but it should be clear that the other syllogisms can be proven similarly. 

\pagebreak

\section{Interface}
\label{sec:2}
Our application, \textit{dasasap}, is motivated by the impact of syllogistic in our education and culture, is inspired upon $\mathcal{L}_{\square}$, and is implemented with LiveCode, which is an event-oriented programming language that allows the development of cross-platform applications.  

At this moment \textit{dasasap} comes with two modes: \textit{arcade} and \textit{learning}. In the arcade mode the user tries to develop valid syllogisms as quick as possible thinking in a jigsaw puzzle style. The objective of this mode is to fasten the recognition of valid syllogisms. The learning mode is a tutorial mode that helps users understand how to develop syllogisms in a jigsaw puzzle style assuming the ideas of $\mathcal{L}_{\square}$ and thus showing the basic concepts of logic (propositions, truth values, form-content distinction, and so on). 

\begin{figure}[h]
  \includegraphics[width=9cm]{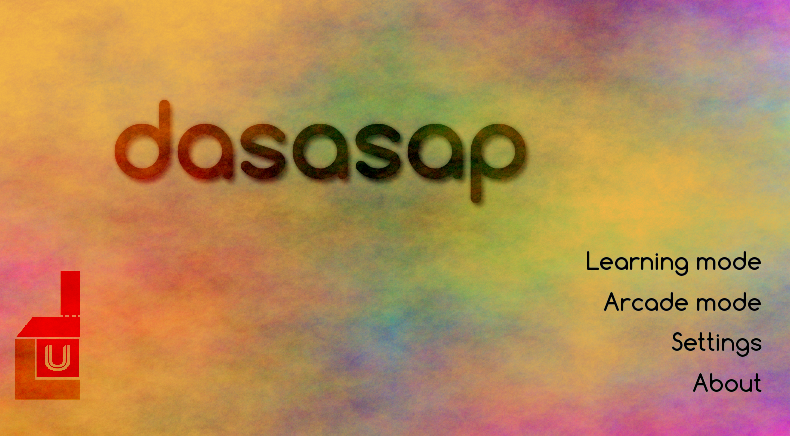}
  \caption{\textit{dasapap}'s home snapshot}  
  \label{fig:11}
\end{figure}

\begin{figure}[h]
  \includegraphics[width=9cm]{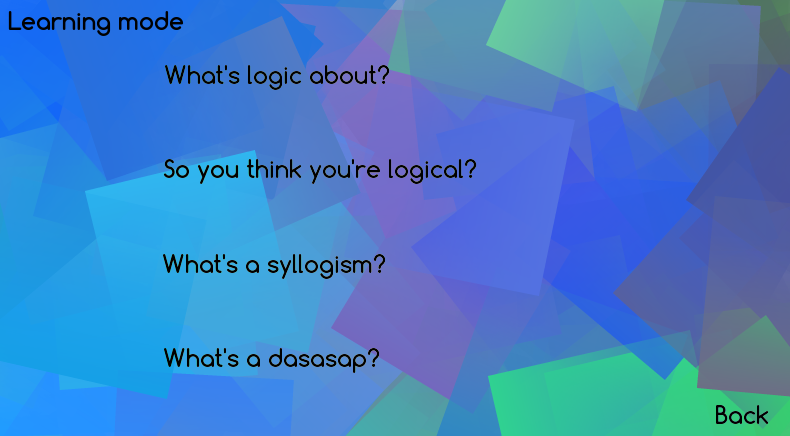}
  \caption{Learning mode snapshot}  
  \label{fig:12}
\end{figure}

\begin{figure}[h]
  \includegraphics[width=9cm]{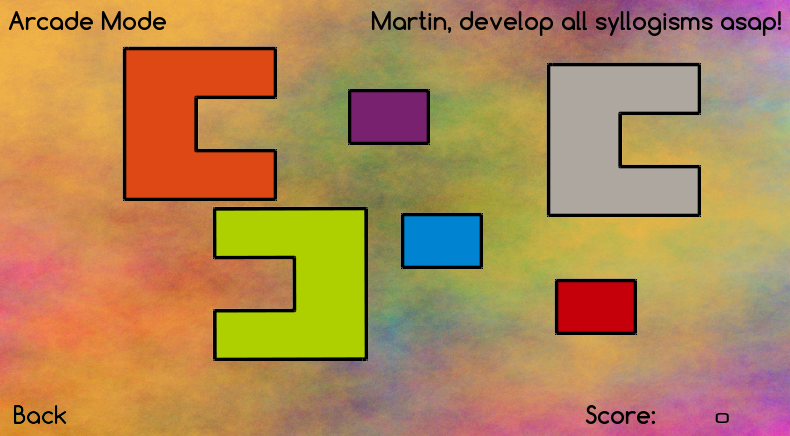}
  \caption{Arcade mode snapshot}  
  \label{fig:13}
\end{figure}

In Figure~\ref{fig:11} we can see the game's home screen running in Ubuntu 12.04. We also depict fragments of the learning and arcade modes (Figures~\ref{fig:12} and~\ref{fig:13}).  
 
The learning mode includes four options. In the option \textit{What's logic about?} the application explains the basic concepts of logic in order to provide an operative definition of logic: proposition, argument, form-content distinction, and  truth-valid distinction are some of the concepts included. The option \textit{So you think you are logical?} leads to a mini-game that produces a ranking: the user is given random syllogisms and the user must decide whether they are valid or invalid. The options \textit{What's a syllogism} and \textit{What's dasasap?} are self-explanatory. 

The arcade mode keeps track of the ranking of the user and allows the construction of syllogisms by measuring time efficiency and accuracy in the development of syllogisms.

\section{Conclusion}
\label{sec:3}
We have introduced an application we call \textit{dasasap} for \textit{develop all syllogisms as soon as possible}. We showed its logical foundations with a system we call $\mathcal{L}_\square$, which is a sound and complete system that uses adequate combinations of polygons as jigsaw puzzles in order to decide the (in)validity of categorical syllogisms. We have implemented the mechanics of this system with LiveCode in order to develop a cross-platform application to aid the teaching-learning process of basic logic. 

Since this application is a small fragment of a larger project we are currently developing, at the moment we are refining the implementation and the overall design. We are also adding a reward system and different match-three puzzle styles.    

\subparagraph*{Acknowledgements}
The authors would like to thank the reviewers for their precise corrections and useful comments. Financial support given by UPAEP Grant 30108-1008.

%





\end{document}